\documentclass[notoc]{JHEP3}
\usepackage{amsfonts,amssymb,amsmath,cite}
\title{Superpotentials for Gauge and Conformal Supergravity Backgrounds}
\author{B. M. Gripaios
\\Department of Physics - Theoretical Physics, University of Oxford, 
\\1, Keble Road, Oxford, OX1 3NP,  UK.% Tel. +44 1865 273982 Fax. +44 1865 273947 
\\ e-mail: \email{b.gripaios1@physics.ox.ac.uk}}
\date{\today}
\abstract{Effective superpotentials obtained by integrating out matter in super Yang-Mills and conformal supergravity backgrounds in $\mathcal{N}=1$ SUSY theories are considered. The pure gauge and supergravity contributions (generalizing Veneziano-Yankielowicz terms) are derived by considering the case with matter fields in the fundamental representation of the gauge group. These contributions represent quantum corrections to the tree-level Yang-Mills and conformal supergravity actions. The classical equations of motion following from the conformal supergravity action require the background to be (super)conformally flat. This condition is unchanged by quantum corrections to the effective superpotential, irrespective of the matter content of the theory.}
\preprint{OUTP-03 31P}
\begin{document}
\section{Introduction}
The effective action of a quantum field theory amounts to a solution of the field theory. Recently, following work by Dijkgraaf and Vafa \cite{Dijkgraaf:2002dh}, a step has been made in this direction for four-dimensional gauge theories with $\mathcal{N}=1$ supersymmetry. Specifically, if one considers the effective action defined by the path integral for an arbitrary gauge theory coupled to matter,\footnote{The theory is not quite arbitrary: the matter must be massive if one is to integrate it out.} then it has been shown how to compute the $F$-terms in the effective action (\emph{i.\ e.\ }the effective superpotential) for a background gauge superfield $W_{\alpha}$. This was shown perturbatively, by computing Feynman diagrams in superspace, in \cite{Dijkgraaf:2002xd} and non-perturbatively, by solving the set of Ward identities following from a generalized Konishi anomaly, in \cite{Cachazo:2002ry}. These arguments determined the effective superpotential up to a pure gauge term (independent of the matter sector couplings), corresponding to the Veneziano-Yankielowicz term \cite{Veneziano:1982ah}. It was subsequently shown \cite{Gripaios:2003kt} that this term could in fact be derived by considering the special case with matter in the fundamental representation and a quartic tree-level matter superpotential.

It was realized concurrently that these methods could be extended to computing the superpotential in a curved supergravity background \cite{Dijkgraaf:2002dh}, specified by an $\mathcal{N}=1$ Weyl superfield $W_{\alpha \beta \gamma}$. This was further developed in refs.\ \cite{Klemm:2002pa,Dijkgraaf:2002yn,Ooguri:2003qp,Ooguri:2003tt,David:2003ke,Alday:2003ms,Ita:2003kk,Dijkgraaf:2003sk}. Yet again, the arguments only determine the superpotential up to pure gauge and supergravity terms (independent of the matter sector couplings).

This paper supplies a direct derivation of the pure gauge and supergravity superpotential using the generalized Konishi anomaly,\footnote{Like the Veneziano-Yankielowicz term, the pure gauge and supergravity superpotential can also be determined from an extended $U(1)_R$ symmetry \cite{Ita:2003kk} or from the measure of a matrix model \cite{Dijkgraaf:2003sk}.} extending the argument of \cite{Gripaios:2003kt}. In the next section, the effective superpotential is computed for matter consisting of $F$ flavours of quarks (and their squark superpartners), transforming in the fundamental representation of the gauge group $SU(N)$,\footnote{The extension to other classical Lie groups is straightforward.} with a quartic tree-level matter superpotential. This matter superpotential allows gauge symmetry breaking via the Higgs mechanism. The computation was originally performed in \cite{Ita:2003kk}. (There is a discrepancy between the effective superpotenital obtained here and that in \cite{Ita:2003kk}, explained in section \ref{SUN}). One proceeds by solving the Ward identities following from the generalized Konishi anomaly. In a flat Minkowski background, correlation functions factorize and all the Ward identities are equivalent. In a supergravity background, the connected parts of two-point correlation functions are non-vanishing \cite{Ita:2003kk} and there are two independent identities. One can show \cite{David:2003ke} that terms of $O((W_{\alpha\beta\gamma}W^{\alpha\beta\gamma})^2)$ vanish in the chiral ring and so functions have a Taylor expansion in the Weyl superfield which terminates at $O(W_{\alpha\beta\gamma}W^{\alpha\beta\gamma})$. Having solved the Ward identities, one can use supersymmetry and holomorphy arguments to determine the effective superpotential up to a `constant' term independent of the matter sector couplings, but depending on the gauge and supergravity background fields and their couplings.

In section \ref{sugra}, the superpotential for fundamental matter is used to derive the superpotential for the pure gauge and supergravity theory. The tree-level matter superpotential allows the gauge group $SU(N)$ to be broken to anything from $SU(N)$ to $SU(N-F)$ at low energy via the Higgs mechanism. One considers two vacua in which the unbroken gauge groups are $SU(N-F_{1,2})$ respectively. Varying the matter couplings, one can take a limit in which the masses of both the quarks and the massive gauge bosons (and their superpartners) become large. In this limit, the massive matter decouples from the unbroken gauge group and its contribution to the effective superpotential can be discarded. What is left must represent the contribution from the unbroken gauge group and the supergravity background. If one subtracts the superpotentials for the two vacua, then the unknown `constants' cancel, resulting in a difference equation whose solution yields the pure gauge and supergravity superpotential. In a suitable renormalization group scheme, the superpotential is
\begin{gather} \label{sup}
W_{\mathrm{eff}} = N \left( -S \log \frac{S}{\Lambda^3}+S\right)
+ \frac{1}{6} (N^2-1) G \log \frac{S}{\Lambda'^3},
\end{gather}
where $S=-\frac{1}{32 \pi^2} \mathrm{tr} W_{\alpha} W^{\alpha}$, $G=\frac{1}{32 \pi^2} \mathrm{tr} W_{\alpha\beta\gamma} W^{\alpha\beta\gamma}$ and $\Lambda$,$\Lambda'$ are dimensional transmutation scales corresponding to the gauge and supergravity couplings.

This superpotential is discussed in section \ref{disc}. The first term (the Veneziano-Yankielowicz superpotential) comes from the tree-level superYang-Mills action, having accounted for quantum corrections. It is shown by analogy that the second term (depending on the Weyl superfield) comes from the (quantum-corrected) tree-level action for conformal supergravity. This is the unique supergravity action which preserves local superconformal symmetry (for a review see \cite{Fradkin:1985am}). It is higher derivative and (in contrast to Einstein-Hilbert gravity) has a dimensionless coupling constant, which has been replaced in (\ref{sup}) by the dimensional transmutation scale $\Lambda'$.

Classically, the equation of motion following from the tree-level conformal supergravity action is $W_{\alpha \beta \gamma}=0$. This implies superconformal flatness, which is not quite trivial (for example, super-anti-de Sitter spacetimes are superconformally flat). Since terms in the effective superpotential can be at most quadratic in $W_{\alpha \beta \gamma}$, it appears that the background Weyl superfield continues to satisfy the superconformal flatness condition irrespective of quantum effects or the actual gauge and matter content of the theory. This may seem paradoxical given that both the gauge and supergravity theories suffer from the conformal anomaly and that the matter sector violates the local superscale invariance even at tree-level if it has dimensionful couplings. However, the claim that conformal flatness persists is only at the level of $F$-terms. It is likely that radiative corrections will generate $D$-terms (\emph{e.\ g.\ } super Einstein-Hilbert terms) which do not respect the local superscale invariance and lead to departures from superconformal flatness.
\section{The Effective Superpotential with Fundamental Quarks}
\label{SUN}
Consider an $\mathcal{N}=1$ supersymmetric theory in four dimensions with matter chiral superfields $\Phi$ coupled to a glueball superfield $S$ and a supergravity superfield given by $G$. The matter superfields $\Phi_I$ transform in some representation of the gauge group denoted by the gauge group index $I$. The quantum theory has the Konishi anomaly \cite{Konishi:1984hf,Konishi:1985tu}; for the generalized chiral rotation $\delta \Phi = \epsilon \Phi' (\Phi)$, one has
\begin{gather} \label{konishi}
\left \langle \Phi'_{I} \frac{\partial W_{\mathrm{tree}} }{\partial \Phi_I} 
 +  \left( 
\frac{1}{32\pi^2} W_{\alpha J}^{K} W^{\alpha J}_{I} + \frac{G}{3} \delta_{I}^{K} \right) \frac{\partial \Phi'_{K}}{\partial \Phi_{I}} \right \rangle = 0,
\end{gather}
where the tree-level matter superpotential $W_{\mathrm{tree}} = g_k \Phi^k$ is some (gauge- and flavour-invariant) polynomial in the matter superfields. The vacuum expectation values are to be evaluated in the presence of a background consisting of the light degrees of freedom. The matter is assumed to be massive and is integrated out. Furthermore, in a generic vacuum of the theory, the gauge symmetry will be spontaneously broken via the Higgs mechanism, with gauge superfields corresponding to broken generators of the gauge group becoming massive. The background should only contain the superfields corresponding to the broken generators (as well as the supergravity superfields).
Let the corresponding background glueball superfield be denoted ${S'}$. The first term in the parentheses in (\ref{konishi}) splits into two parts. The part tracing over the unbroken gauge group is by definition ${S'}$. The other part traces over the broken part of the gauge group. The associated superfields are massive, and their potential is (classically) quadratic and centred at the origin, such that their vevs are zero.\footnote{This is certainly true at the classical level, but it is possible that quantum corrections will modify this. However, the limit will be taken later later on in which the masses of the gauge bosons go to infinity and they decouple. In this limit, their vevs certainly are zero, and so the possibility of quantum corrections will not affect the argument below.}

The effective superpotential for the background superfields can then be determined by solving the partial differential equations
\begin{gather} \label{pdes}
\frac{\partial W_{\mathrm{eff}} }{\partial g_k} = \langle \Phi^k \rangle,
\end{gather}
which follow by holomorphy and supersymmetry. These determine $W_{\mathrm{eff}}$ up to a constant term independent of the matter couplings $g_k$.

Now consider the particular case where the matter sector consists of $F$ flavours of `quarks', \emph{viz.\ }$F$ chiral superfields $Q_{I}^{i}$ transforming in the fundamental representation of the gauge group and $F$ chiral superfields $\tilde{Q}^{J}_{j}$ in the anti-fundamental representation, where $i$ and $j$ are flavour indices. The tree-level matter superpotential (for $F<N$) is written in terms of the $F \times F$ gauge-invariant meson matrix $M_{j}^{i}=Q_{I}^{i}\tilde{Q}^{I}_{j}$ as
\begin{gather} \label{wtree}
W_{\mathrm{tree}} = m \mathrm{tr} M - \lambda \mathrm{tr} M^2.
\end{gather}
The classical equations of motion for the matter fields are
\begin{gather} \label{classeom}
mM_{i}^{j} - 2\lambda M_{i}^{k}M_{k}^{j} = 0.
\end{gather}
The meson matrix $M$ can be brought to diagonal form via a global $SU(F)$ flavour transformation; the classical vacua then have $F_{-}$ eigenvalues at $M_{i}^{i}=0$ and $F_{+}=F-F_{-}$ eigenvalues at  $M_{i}^{i}= m/2\lambda$ (no sum on $i$), with the low energy gauge group broken down to $SU(N-F_{+})$. 

Turning to the generalized Konishi anomaly (\ref{konishi}), the variations
\begin{align}
\delta Q_{I}^{f} &= \epsilon Q_{I}^{f'}, \nonumber \\
\delta Q_{I}^{f} &= \epsilon Q_{I}^{f'} M_{h}^{h'}, \nonumber \\
&\phantom{=}\vdots
\end{align}
yield anomalous Ward identities for the meson matrix
\begin{align} \label{quanteom}
m \langle M_{f}^{f'} \rangle - 2 \lambda \langle (M^2)_{f}^{f'} \rangle &= \delta_{f}^{f'} ({S'} - \frac{N}{3} G), \nonumber \\
m \langle M_{f}^{f'} M_{h}^{h'}\rangle - 2 \lambda \langle (M^2)_{f}^{f'} M_{h}^{h'} \rangle &= \delta_{f}^{f'} ({S'} - \frac{N}{3} G)\langle  M_{h}^{h'} \rangle - \frac{1}{3} G \delta_{f}^{h'} \langle M_{h}^{f'}\rangle, \nonumber \\
&\phantom{=}\vdots
\end{align}
The aim is to solve the complete set of such Ward identities for the matter expectation values. Since terms of $O(G^2)$ or higher vanish in the chiral ring (and \emph{ergo} in vacuum expectation values) it is convenient to Taylor expand everywhere in powers of $G$ and to perform the analysis term by term. 

First recall the analysis at $O(G^0)$ performed previously \cite{Brandhuber:2003va,Gripaios:2003kt}, corresponding to a flat background. At this order, correlation functions factorize, because connected $n$-point functions are of $O(G^{n})$ \cite{Ita:2003kk}. The Ward identities (\ref{quanteom}) all reduce to the single identity
\begin{gather} \label{}
m \langle M_{i}^{j} \rangle  - 2\lambda \langle M_{i}^{k} \rangle \langle M_{k}^{j} \rangle = \delta^{j}_{i} {S'}.
\end{gather}
There is only one independent \emph{vev}: the expectation value of the meson matrix. This can be diagonalized by a global $SU(F)$ rotation in flavour space. In such a basis, the Ward identities have the solution
\begin{gather} \label{}
\langle M_{i}^{j}\rangle =  \delta_{i}^{j} \frac{m}{4\lambda}\left( 1 \pm \sqrt{1-\frac{8\lambda {S'}}{m^2}} \right).
\end{gather}
The quantum vacua have $F_{\pm}$ eigenvalues at each of these two values corresponding to Higgsed or un-Higgsed quarks. 

At  $O(G^1)$, things become more complicated, because the connected two-point correlation function is not zero. One must consider matrix \emph{vevs} which are the \emph{vevs} of products of the meson matrix. The global flavour symmetry can be used to diagonalize one such matrix \emph{vev}, chosen to be $\langle M_{i}^{j} \rangle$ as before. Since there is still a residual $SU(F_{+}) \times SU(F_{-})$ global flavour symmetry, the \emph{vev} of the meson matrix must take the form
\begin{gather} \label{M}
\langle  M_{\pm f}^{\phantom{\pm} f'} \rangle = M_{\pm} \delta _{f}^{f'}.
\end{gather}
(The subscript $\pm$ labels the Higgsed and un-Higgsed spaces, to conform with the notation below.)
It is claimed furthermore that all matrix \emph{vevs} are \emph{block}-diagonal in this basis. That is, they act reducibly on the Higgsed and un-Higgsed flavour subspaces of dimensions  $F_{+}$ and  $F_{-}$ respectively.\footnote{This claim is not an assumption, but an \emph{ansatz}: one is trying to solve the set of Ward identities (\ref{quanteom}); it will be seen that the block-diagonalization does this.}\footnote{The work \cite{Ita:2003kk} does not make this \emph{ansatz}, but rather starts from the \emph{ans\"{a}tze} 
$\langle  M_{ f}^{ f'} \rangle = M \delta _{f}^{f'}$ and 
$\langle  M_{ f}^{ f'}  M_{ g}^{ g'} \rangle_c =
\delta _{f}^{g'} \delta _{g}^{f'}  B$; 
this still yields two values for $M$ and $B$, but the solution is only valid if all the eigenvalues of $\langle  M_{ f}^{ f'} \rangle$ are the same. This is true for the cases $F=F_+$ and $F=F_-$, but not in the general case.} The Konishi anomaly equations (\ref{quanteom}) are valid \emph{verbatim} in each subspace separately, with the flavour indices running from $1$ to $F_{+}$ in the Higgsed subspace (labelled by a plus sign as in (\ref{M}) and from  $1$ to $F_{-}$ in the un-Higgsed subspace (labelled by a minus).

The two-point function no longer factorises, but has a connected part of $O(G)$
\begin{gather}
\langle  M_{\pm f}^{\phantom{\pm} f'}  M_{\pm g}^{\phantom{\pm} g'} \rangle =
\langle  M_{\pm f}^{\phantom{\pm} f'}  M_{\pm g}^{\phantom{\pm} g'} \rangle_c
+ M^{2}_{\pm} \delta _{f}^{f'} \delta _{g}^{g'}
\end{gather}
and one makes the further \emph{ansatz} that
\begin{gather} \label{Mtwo}
\langle  M_{\pm f}^{\phantom{\pm} f'}  M_{\pm g}^{\phantom{\pm} g'} \rangle_c =
\delta _{f}^{g'} \delta _{g}^{f'}  B_{\pm} .
\end{gather}
The connected part of the three-point function vanishes to $O(G)$, leaving
\begin{gather}
\langle  M_{\pm f}^{\phantom{\pm} f'}  M_{\pm g}^{\phantom{\pm} g'}  M_{\pm h}^{\phantom{\pm} h'}\rangle=
 M^{3}_{\pm}  \delta _{f}^{f'} \delta _{g}^{g'} \delta _{h}^{h'} + 
\langle  M_{\pm f}^{\phantom{\pm} f'}  M_{\pm g}^{\phantom{\pm} g'} \rangle_c M_{\pm} \delta _{h}^{h'} + \mathrm{permutations},
\end{gather}
such that 
\begin{gather}
\langle ( M^{2}_{\pm})_{f}^{ f'}  M_{\pm h}^{\phantom{\pm} h'} \rangle = \delta _{f}^{f'} \delta _{h}^{h'} (F_{\pm} B_{\pm} M_{\pm} +M^{3}_{\pm})
+ \delta _{f}^{h'} \delta _{h}^{f'} 2  B_{\pm} M_{\pm}.
\end{gather}
Taylor expanding in $G$, there are no contributions beyond $O(G)$; from (\ref{Mtwo}), $B_{\pm}$ is already of $O(G)$ and one can write
\begin{align} \label{comps}
M_{\pm} &= M_{0\pm} + M_{1\pm} G, \nonumber \\
B_{\pm} &= B_{1\pm} G.
\end{align}
The set of Ward identities (\ref{quanteom}) reduce to three equations in three unknowns, \emph{viz.}
\begin{align}
{S'} &= m M_{0\pm} - 2 \lambda M^{2}_{0\pm}, \nonumber \\
0 &= m M_{1\pm} - 2 \lambda (2M_{0\pm} M_{1\pm} + F_{\pm} B_{1\pm}) + \frac{N}{3}, \nonumber \\
0 &= m  B_{1\pm} - 4 \lambda  B_{1\pm} M_{0\pm} + \frac{M_{0\pm}}{3},
\end{align}
with the solutions
\begin{align} \label{vevs}
M_{0\pm} &= \frac{m}{4 \lambda} \left( 1 \pm \sqrt{1-\frac{8 \lambda {S'}}{m^2}} \right), \nonumber \\
M_{1\pm} &= \pm \frac{(2N-F_{\pm})}{6m\sqrt{1-\frac{8 \lambda {S'}}{m^2}}} - \frac{F_{\pm}}{6m(1-\frac{8 \lambda {S'}}{m^2})},\nonumber \\
B_{1\pm} &= \frac{1}{12 \lambda} \left( 1 \pm \frac{1}{\sqrt{1-\frac{8 \lambda {S'}}{m^2}}} \right). 
\end{align}
The partial differential equations (\ref{pdes}) for the effective superpotential with respect to the matter sector couplings are
\begin{align} \label{pdes2}
\frac{\partial W_{\mathrm{eff}}}{\partial m} &= \langle \mathrm{tr} M \rangle, \nonumber \\
\frac{\partial W_{\mathrm{eff}}}{\partial \lambda} &= - \langle \mathrm{tr} M^2 \rangle.
\end{align}
Again it is convenient to Taylor expand $ W_{\mathrm{eff}}$ in $G$,
\begin{gather}
 W_{\mathrm{eff}} = W_{0\mathrm{eff}} + W_{1\mathrm{eff}} G.
\end{gather}
At $O(G^0)$, one has \cite{Brandhuber:2003va,Gripaios:2003kt}
\begin{multline} \label{wzero}
W_{0\mathrm{eff}} 
= F\frac{m^2}{8 \lambda} + (F_{+}-F_{-})\frac{m^2}{8 \lambda}\sqrt{1-\frac{8\lambda {S'}}{m^2}}
+ F {S'} \log m \\
+ {S'} \log \left[ 
\left( \frac{1}{2} +  \frac{1}{2}\sqrt{1-\frac{8\lambda {S'}}{m^2}} \right)^{F_{-}} 
\left( \frac{1}{2} -  \frac{1}{2}\sqrt{1-\frac{8\lambda {S'}}{m^2}} \right)^{F_{+}} 
\right]
+ c({S'}),
\end{multline}
where $c({S'})$ is independent of $m$ and $\lambda$.
At $O(G^1)$, one has, from (\ref{comps}) and (\ref{pdes2}), the partial differential equations
\begin{align}
\frac{\partial W_{1\mathrm{eff}}}{\partial m} &= F_{+}  M_{1+} +  F_{-}  M_{1-}, \nonumber \\
\frac{\partial W_{1\mathrm{eff}}}{\partial \lambda} &= - F_{+} (F_{+} B_{+} + 2 M_{0+} M_{1+}) - F_{-} (F_{-} B_{-} + 2 M_{0-} M_{1-}).
\end{align}
Substituting from (\ref{vevs}), one obtains the solution
\begin{align} \label{wtwo}
W_{1\mathrm{eff}} = &- F_{+} \left( \frac{N}{3} \log m
+ \frac{2N - F_{+}}{6} \log \left[ \frac{1}{2} - \frac{1}{2} \sqrt{1-\frac{8\lambda {S'}}{m^2}} \right] + \frac{F_{+}}{12} \log \left[ 1-\frac{8\lambda {S'}}{m^2} \right] \right) \nonumber \\
&- F_{-} \left( \frac{N}{3} \log m
+ \frac{2N - F_{-}}{6} \log \left[ \frac{1}{2} + \frac{1}{2} \sqrt{1-\frac{8\lambda {S'}}{m^2}} \right] + \frac{F_{-}}{12} \log \left[ 1-\frac{8\lambda {S'}}{m^2} \right] \right) \nonumber \\
&+ d({S'}),
\end{align}
where $d({S'})$ is independent of $m$ and $\lambda$ and is determined in the next section.

\section{The Pure Gauge and Supergravity Superpotential} \label{sugra}
The pure gauge and supergravity superpotential can be determined as follows. First take the limit in which both the quark masses, $m$, and the gauge boson masses, $\sqrt{m/ 2 \lambda}$, become large. The effective superpotential (\ref{wtwo}) becomes
\begin{gather} \label{deltaw2}
 W_{1\mathrm{eff}} \stackrel{m^2/\lambda \rightarrow \infty}{\rightarrow}-\frac{NF}{3}\log m - F_+ \frac{2N-F_+}{6} \log \frac{2 \lambda {S'}}{m^2}. 
\end{gather}
What is the meaning of this expression? In this limit, the massive degrees of freedom decouple from the pure $SU(N-F_{+})$ gauge theory in the supergravity background. The effective superpotential should consist of the superpotential for the low energy pure gauge and supergravity theory plus terms representing the contribution of the decoupled matter. The decoupled matter consists of the quarks and the massive gauge bosons corresponding to the broken generators of $SU(N)$ (and their superpartners). The quark superfields have been integrated out, and one can calculate their contribution to the effective superpotential as follows. The non-renormalization theorem applies and the contribution is found by replacing the quark fields in the tree-level superpotential (\ref{wtree}) by their classical vacuum expectation values. This gives a contribution to $W_{0\mathrm{eff}}$ and was discussed in \cite{Gripaios:2003kt}. The contribution of the massive gauge bosons was earlier seen to be zero. Thus one sees that the terms in (\ref{deltaw2}) do indeed correspond to the contribution from the low energyunbroken pure gauge theory and supergravity. This contains the unknown constant $d$, which can be removed by considering the superpotentials for two distinct vacua in which the number of Higgsed quarks, $F_+$, takes the values $F_1$ and $F_2$, but the argument ${S'}$ takes the same value, $T$ say, in both.\footnote{Of course the physical interpretation of $T$ is different in the two vacua. Here however one simply wants to determine the functional form of $W_{1\mathrm{eff}}$. $W_{1\mathrm{eff}}$ is an unconstrained function of its arguments and so the arguments may be chosen arbitrarily.} If one then subtracts the two effective superpotential functions, the unknown constant $d$ cancels, giving 
\begin{gather} \label{dwgauge}
\Delta W_{1\mathrm{eff}} =
- F_1 \frac{2N-F_1}{6} \log \frac{2 \lambda {T}}{m^2}
+ F_2 \frac{2N-F_2}{6} \log \frac{2 \lambda {T}}{m^2}
\end{gather}
This expression ought to involve only the low energy degrees of freedom and the gauge and  gravitational couplings, yet it appears to depend on the tree-level matter couplings $m$ and $\lambda$. The same phenomenon was observed with the pure gauge theory superpotential in \cite{Gripaios:2003kt} and was shown to be due to the requirement of scale matching: the running gauge couplings of the Higgsed and un-Higgsed theories (which have different beta functions) must match at the Higgs scale, and this leads to a relation between the strong coupling scales of the high and low energy theories. In the gravitational case, the same thing happens: loop diagrams result in logarithmic corrections to the supergravity coupling which depend on the fields running in the loops and so there is again a scale-matching constraint for the Higgsed and un-Higgsed theories. The beta function can be found from the trace or conformal anomaly \cite{Christensen:1978gi,McArthur:1984fk}. The one-loop coefficient is given by
\begin{gather}
b = - 3 N_v + N_{\chi},
\end{gather} 
where $N_v$ and $N_{\chi}$ are the number of vector and chiral SUSY multiplets respectively. The scale matching must be performed at two intermediate scales, corresponding to the quark mass, $m$, and the gauge boson mass, $\sqrt{m/2 \lambda }$. At high energies, one has $N^2-1$ massless vector supermultiplets and $2 N F $ chiral supermultiplets. Below the Higgs scale, there are $(N-F_+)^2-1$ vector supermultiplets and $2 N F_- + F_{+}^{2} $ chiral supermultiplets,\footnote{To see this, note that each of the $2 N F_+ - F_{+}^{2}$ massless vector supermultiplets corresponding to broken generators eats a chiral supermultiplet to become a massive vector supermultiplet, leaving $2 N F_- + F_{+}^{2} $ of the original $2NF$ chiral supermultiplets.} such that one has
\begin{gather}
\left( \frac{{\Lambda'}_{N,F}}{\sqrt{m/2 \lambda }} \right)^{-3(N^2-1) + 2NF} = 
\left( \frac{{\Lambda'}_{N-F_+,F_-}}{\sqrt{m/2 \lambda }} \right)^{-3 [(N-F_+)^2-1]+2 N F_- + F_{+}^{2}}.
\end{gather}
Here, ${\Lambda'}_{N,F}$ is the dimensional transmutation scale corresponding to the supergravity coupling of the theory with gauge group $SU(N)$ coupled to $F$ quark flavours. Below the quark mass scale one has only  $(N-F_+)^2-1$ vector supermultiplets, such that 
\begin{gather}
\left( \frac{{\Lambda'}_{N-F_+,F_-}}{m} \right)^{-3 [(N-F_+)^2-1]+2 N F_- + F_{+}^{2}} = 
\left( \frac{{\Lambda'}_{N-F_+,0}}{m} \right)^{-3 [(N-F_+)^2-1]}.
\end{gather}
The scale matching relation is therefore
\begin{gather} \label{sugrascale}
{\Lambda'}_{N-F_{+},0}^{3((N-F_{+})^2-1)} \left( \frac{m^2}{2\lambda} \right)^{2NF_{+}- F_{+}^{2}}
 = {\Lambda'}_{N,F}^{3(N^2-1) - 2 N F} m^{2NF}.
\end{gather}
Using this relation, one can replace the matter sector couplings $m$ and $\lambda$ in (\ref{deltaw2}) by the appropriate gravitational scales to obtain
\begin{multline} \label{}
W_{1,\mathrm{eff}}(N-F_{1},T,{\Lambda'}_{N-F_{1},0}) - W_{1,\mathrm{eff}}(N-F_{2},T,\Lambda'_{N-F_{2},0}) 
= \\ \frac{1}{6}((N-F_{1})^2-1)  \log \frac{T}{{\Lambda'}_{N-F_{1},0}^{3}}
-\frac{1}{6}((N-F_{2})^2-1)  \log \frac{T}{{\Lambda'}_{N-F_{2},0}^{3}},
\end{multline}
where the functional dependence of $W_{1\mathrm{eff}}$ has been indicated explicitly. This difference equation has the solution
\begin{gather} \label{}
W_{1,\mathrm{eff}}(N,S,{\Lambda'}_{N,0}) = 
\frac{1}{6}(N^2-1) \log \frac{S}{{\Lambda'}_{N,0}^{3}} + g(S), 
\end{gather}
where $S$ has been re-instated and $g(S)$ is an arbitrary function of $S$ alone: it cannot depend on any of the other parameters. Furthermore, $gG$ (which appears in the superpotential) must be of dimension three and $g$ must therefore be of dimension zero, \emph{i.\ e.\ }a pure number. This ambiguity in $W_{1\mathrm{eff}}$ corresponds to a re-scaling of ${\Lambda'}^{N^2-1}$, or equivalently to the freedom to choose a renormalisation group scheme.
In a scheme in which $g$ vanishes, the complete effective superpotential for the pure $SU(N)$ gauge and supergravity theory is
\begin{gather} \label{sunw}
W_{\mathrm{eff}}(N,S,\Lambda_{N,0}) = 
N \left( -S \log \frac{S}{\Lambda^3} + S \right)
+ \frac{1}{6}(N^2-1) G \log \frac{S}{{\Lambda'}^{3}},
\end{gather}
which is the form obtained by extended $U(1)_R$ symmetry considerations \cite{Ita:2003kk}. There it was noted that there is no need for the scales $\Lambda$ and $\Lambda'$ to be the same to satisfy the $U(1)_R$ symmetry. The above derivation shows that these scales are indeed distinct: they represent the gauge and gravitational couplings.

Now that the effective superpotential for the low energy gauge theory has been derived, one can determine $d(S)$ in (\ref{wtwo}) as in \cite{Brandhuber:2003va} by demanding that $W_{1\mathrm{eff}}$ in (\ref{wtwo}) reproduces the correct limit as $m^2/\lambda \rightarrow \infty$ for the vacuum with $F_{+}$ Higgsed quarks and low energy gauge group $SU(N-F_{+})$. One obtains
\begin{align} \label{wthree}
W_2 = &- F_+ \left( 
+ \frac{2N - F_+}{6} \log \left[ \frac{1}{2} - \frac{1}{2} \sqrt{1-\frac{8\lambda S}{m^2}} \right] + \frac{F_+}{12} \log \left[ 1-\frac{8\lambda S}{m^2} \right] \right) \nonumber \\
&- F_- \left( 
+ \frac{2N - F_-}{6} \log \left[ \frac{1}{2} + \frac{1}{2} \sqrt{1-\frac{8\lambda S}{m^2}} \right] + \frac{F_-}{12} \log \left[ 1-\frac{8\lambda S}{m^2} \right] \right) \nonumber \\
&+ \frac{1}{6} \log \frac{S^{N^2-1}}{{\Lambda'}^{3(N^2-1)-2NF}m^{2NF}}.
\end{align} 
The constant $d$ is seen to depend only on the glueball and supergravity backgrounds and their couplings, justifying \emph{a posteriori} the assumption that it cancels in the vacuum subtraction.
\section{Discussion}
\label{disc}
Having derived the form of the effective superpotential for the pure gauge and supergravity theory (\ref{sunw}), it is of interest to study the vacuum structure obtained upon minimisation. Before doing so, it is important to clarify what is meant by `the supergravity theory', since there exists more than one. To do this, one can proceed by analogy with the pure gauge theory case, with a flat background and no matter. There one starts with the tree-level Yang-Mills superpotential $2\pi i \tau S$ (plus Hermitian conjugate) which is renormalized to $N \log \Lambda^3 S$. 
The non-perturbative dynamics modifies this further to
\begin{gather}
N \left( -S \log \frac{S}{\Lambda^3} + S \right).
\end{gather}
The effective potential is thus generated in a natural way from the tree-level Yang-Mills superpotential. In a supergravity background, a term
\begin{gather}
\frac{1}{6}(N^2-1)  \log \frac{S}{{\Lambda'}^{3}} G
\end{gather}
is added to the superpotential. Running the previous argument backwards, it seems reasonable that this comes from the renormalized superpotential
\begin{gather}
-\frac{1}{6}(N^2-1)  \log {\Lambda'}^{3} G,
\end{gather}
which in turn comes from the bare, tree-level superpotential $2\pi i \tau' G$, where $\tau'$ is some dimensionless coupling.
Is this a \emph{bona fide} supergravity action? Indeed it is, being (a gauge-fixed version of) the unique action preserving the group of local superscale transfromations and known as \emph{conformal supergravity} (for a review see \cite{Fradkin:1985am}).\footnote{Conventional supergravities extending the Einstein-Hilbert action only preserve the subgroup of local super Poincar\'{e} transformations.} It is higher-derivative and has a logarithmically-renormalized dimensionless coupling constant. 

This suggests that the effective superpotentials (\ref{sunw}) and (\ref{wthree}) encode quantum corrections to conformal supergravity; if this is the case, what are the implications for the vacuum structure? Classically, the equation of motion for the Weyl superfield following from the conformal supergravity action is $W_{\alpha\beta\gamma}=0$. This is not quite trivial: it does not imply Minkowski flatness, but rather (super)conformal flatness. In particular, super anti-de Sitter spacetimes with a cosmological constant are permitted.\footnote{But not super-de Sitter: there are no SUSY extensions of the de Sitter algebra.} When gauge and matter supermultiplets are added, along with their quantum effects, the effective superpotential contains terms of $O(G^0)$ and $O(G)$. The former must be present even in a Minkowski background and so the latter represent the effects of a curved background in their entirety. Since $G$ is quadratic in the Weyl superfield, the classical equation of motion $W_{\alpha\beta\gamma}=0$ is unchanged. Thus superconformal flatness is preserved irrespective of the the gauge and matter content of the theory. If this were the whole story, it would be rather remarkable, since the addition of gauge fields, matter and quantum effects leads to the breaking of the local superscale invariance via the superconformal anomaly, even if the matter sector is superscale-invariant, which it need not be.

There is, however, an important \emph{caveat}. The above arguments only allow one to determine the effective superpotential, that is the $F$-terms in the effective action. Just as for the theory with global supersymmetry, one is unable to make general statements about the $D$-terms in the effective action. Since the superscale invariance is broken (either explicitly through dimensionful couplings in the tree-level matter superpotential, or via the superconformal anomaly), there is nothing to prevent the generation of non-scale-invariant gravitational corrections (such as super Einstein-Hilbert terms) via quantum effects, even if one starts with classical conformal supergravity. These corrections appear in the effective action as $D$-terms, and would in general affect the vacuum structure.
\acknowledgments
The author would like to thank J. March-Russell for discussions and 
is supported by a PPARC studentship.

\providecommand{\href}[2]{#2}\begingroup\raggedright\endgroup


\begin{thebibliography}{10}

\bibitem{Dijkgraaf:2002dh}
R.~Dijkgraaf and C.~Vafa, {\it A perturbative window into non-perturbative
  physics},  \href{http://xxx.lanl.gov/abs/hep-th/0208048}{{\tt
  hep-th/0208048}}.

\bibitem{Dijkgraaf:2002xd}
R.~Dijkgraaf, M.~T. Grisaru, C.~S. Lam, C.~Vafa, and D.~Zanon, {\it
  Perturbative computation of glueball superpotentials},
  \href{http://xxx.lanl.gov/abs/hep-th/0211017}{{\tt hep-th/0211017}}.

\bibitem{Cachazo:2002ry}
F.~Cachazo, M.~R. Douglas, N.~Seiberg, and E.~Witten, {\it Chiral rings and
  anomalies in supersymmetric gauge theory},  {\em JHEP} {\bf 12} (2002) 071,
  [\href{http://xxx.lanl.gov/abs/hep-th/0211170}{{\tt hep-th/0211170}}].

\bibitem{Veneziano:1982ah}
G.~Veneziano and S.~Yankielowicz, {\it An effective Lagrangian for the pure N=1
  supersymmetric Yang-Mills theory},  {\em Phys. Lett.} {\bf B113} (1982) 231.

\bibitem{Gripaios:2003kt}
B.~M. Gripaios and J.~F. Wheater, {\it Veneziano-Yankielowicz superpotential
  terms in N = 1 SUSY gauge theories},
  \href{http://xxx.lanl.gov/abs/hep-th/0307176}{{\tt hep-th/0307176}}.

\bibitem{Klemm:2002pa}
A.~Klemm, M.~Marino, and S.~Theisen, {\it Gravitational corrections in
  supersymmetric gauge theory and matrix models},  {\em JHEP} {\bf 03} (2003)
  051, [\href{http://xxx.lanl.gov/abs/hep-th/0211216}{{\tt hep-th/0211216}}].

\bibitem{Dijkgraaf:2002yn}
R.~Dijkgraaf, A.~Sinkovics, and M.~Temurhan, {\it Matrix models and
  gravitational corrections},
  \href{http://xxx.lanl.gov/abs/hep-th/0211241}{{\tt hep-th/0211241}}.

\bibitem{Ooguri:2003qp}
H.~Ooguri and C.~Vafa, {\it The C-deformation of gluino and non-planar
  diagrams},  \href{http://xxx.lanl.gov/abs/hep-th/0302109}{{\tt
  hep-th/0302109}}.

\bibitem{Ooguri:2003tt}
H.~Ooguri and C.~Vafa, {\it Gravity induced C-deformation},
  \href{http://xxx.lanl.gov/abs/hep-th/0303063}{{\tt hep-th/0303063}}.

\bibitem{David:2003ke}
J.~R. David, E.~Gava, and K.~S. Narain, {\it Konishi anomaly approach to
  gravitational F-terms},  \href{http://xxx.lanl.gov/abs/hep-th/0304227}{{\tt
  hep-th/0304227}}.

\bibitem{Alday:2003ms}
L.~F. Alday, M.~Cirafici, J.~R. David, E.~Gava, and K.~S. Narain, {\it
  Gravitational F-terms through anomaly equations and deformed chiral rings},
  \href{http://xxx.lanl.gov/abs/hep-th/0305217}{{\tt hep-th/0305217}}.

\bibitem{Ita:2003kk}
H.~Ita, H.~Nieder, and Y.~Oz, {\it Gravitational F-terms of N = 1
  supersymmetric SU(N) gauge theories},
  \href{http://xxx.lanl.gov/abs/hep-th/0309041}{{\tt hep-th/0309041}}.

\bibitem{Dijkgraaf:2003sk}
R.~Dijkgraaf, M.~T. Grisaru, H.~Ooguri, C.~Vafa, and D.~Zanon, {\it Planar
  gravitational corrections for supersymmetric gauge theories},
  \href{http://xxx.lanl.gov/abs/hep-th/0310061}{{\tt hep-th/0310061}}.

\bibitem{Fradkin:1985am}
E.~S. Fradkin and A.~A. Tseytlin, {\it Conformal supergravity},  {\em Phys.
  Rept.} {\bf 119} (1985) 233--362.

\bibitem{Konishi:1984hf}
K.~Konishi, {\it Anomalous supersymmetry transformation of some composite
  operators in SQCD},  {\em Phys. Lett.} {\bf B135} (1984) 439.

\bibitem{Konishi:1985tu}
K.-I. Konishi and K.-I. Shizuya, {\it Functional integral approach to chiral
  anomalies in supersymmetric gauge theories},  {\em Nuovo Cim.} {\bf A90}
  (1985) 111.

\bibitem{Brandhuber:2003va}
A.~Brandhuber, H.~Ita, H.~Nieder, Y.~Oz, and C.~Romelsberger, {\it Chiral
  rings, superpotentials and the vacuum structure of N = 1 supersymmetric gauge
  theories},  \href{http://xxx.lanl.gov/abs/hep-th/0303001}{{\tt
  hep-th/0303001}}.

\bibitem{Christensen:1978gi}
S.~M. Christensen and M.~J. Duff, {\it Axial and conformal anomalies for
  arbitrary spin in gravity and supergravity},  {\em Phys. Lett.} {\bf B76}
  (1978) 571.

\bibitem{McArthur:1984fk}
I.~N. McArthur, {\it Super b(4) coefficients in supergravity},  {\em Class.
  Quant. Grav.} {\bf 1} (1984) 245.

\end{thebibliography}
\end{document}